\documentclass[aps,nofootinbib,showpacs]{revtex4}
\begin{document}

\rightline{Feb 2004}

\title{Explaining $\Omega_{Baryon} \approx 0.2\, \Omega_{Dark}$ through the
synthesis of ordinary matter from mirror matter: a more general analysis}

\author{R. Foot}\email{foot@physics.unimelb.edu.au}
\author{R. R. Volkas}\email{r.volkas@physics.unimelb.edu.au}
\affiliation{School of Physics, Research Centre for High Energy Physics,
The University of Melbourne, Victoria 3010, Australia}

\begin{abstract}
The emerging cosmological picture
is of a spatially flat universe composed predominantly of three components:
ordinary baryons ($\Omega_B \approx 0.05$), non-baryonic 
dark matter ($\Omega_{Dark} \approx 0.22$) and dark energy
($\Omega_{\Lambda} \approx 0.7$). We recently proposed that ordinary
matter was synthesised from mirror matter, motivated by the argument that the
observed similarity of $\Omega_B$ and $\Omega_{Dark}$
suggests an underlying similarity between the fundamental properties of
ordinary and dark matter particles. In this paper we generalise the
previous analysis by considering a wider class of effective operators
that non-gravitationally couple the ordinary and mirror sectors.
We find that while all considered operators imply
$\Omega_{Dark} = $ few$\times \Omega_B$, only a subset quantitatively
reproduce the observed ratio $\Omega_B/\Omega_{Dark} \approx 0.20$.
The $\sim 1$ eV mass scale induced through these operators hints
at a connection with neutrino oscillation physics.
\end{abstract}

\maketitle

\section{Introduction}

A variety of evidence, culminating in the Wilkinson Microwave Anistropy Probe 
measurements of the cosmic microwave background \cite{cmb}, points to a spatially
flat universe composed of approximately $5\%$ baryons (B), $22\%$ 
non-baryonic dark matter (DM) and roughly $70\%$ dark energy. These
fractions, while different, are suspiciously similar in magnitude.
From a theoretical point of view, this is surprising, as 
one might {\it a priori} expect the physics of each component to be rather 
different. This is a new naturalness puzzle.

The similarity of the positive and negative pressure components, roughly
$30\%$ and $70\%$ respectively, has received some attention in the literature.
However, the similar magnitudes of $\Omega_B$ and $\Omega_{Dark}$ are also
puzzling. (As usual, $\Omega_X$ denotes the ratio of the
energy density of component $X$ and the critical density.)
In Ref.\ \cite{paper1}, we began an exploration of a possible solution. We
proposed that the similar baryonic and dark matter densities suggest that
the internal microphysics of each component are actually similar or 
identical.
Specifically, this will be case if the dark matter is identified with
mirror baryonic matter, $B'$.\footnote{The
$\Omega_B \sim \Omega_{Dark}$ issue was very briefly touched upon by Bento and Berezhiani
\cite{bento} also in the context of mirror matter models. Note, though, that their mechanism 
produces $\Omega_B = \Omega'_B$ for the symmetric mirror model.} 
The purpose of this new paper is to extend the analysis
of Ref.\ \cite{paper1}.

A theoretical motivation for mirror matter \cite{parity,parity2}
is to retain the full Poincar\'{e}
group, including improper Lorentz transformations such as parity inversion
and time reversal, as an exact symmetry group of nature, despite the $V-A$
character of weak interactions. Mirror matter theories 
have a $G \otimes G$ gauge
group structure, with ordinary and mirror particles assigned to $(R,1)$ 
and $(1,R)$ representations, respectively. Under parity transformations,
a given ordinary particle interchanges with its mirror partner. For fermions,
a left-handed (right-handed) ordinary particle is paired with a
right-handed (left-handed) mirror partner. Time-reversal invariance $T'$
follows from the $CPT$ theorem, with $T' P' \equiv CPT$ where $P'$ is
the exact mirror parity operation defined above. We shall focus on the
simplest mirror matter model, where $G$ is the standard model gauge group,
there are no exotic fermions besides the mirror partners of known
quarks and leptons, and there is just one Higgs doublet paired with one
mirror Higgs doublet. We shall consider only the case where the vacuum
also respects improper Lorentz transformations, that is, the discrete
spacetime symmetries are not spontaneously broken \cite{parity2}. This means that
an ordinary particle and its mirror partner have the same mass. A large region
of Higgs potential parameter space allows this aesthetically appealing
scenario \cite{parity2}.

Mirror particles interact amongst themselves through their own
versions of the strong, weak and electromagnetic interactions. The
two sectors are inevitably coupled by gravity, but various
non-gravitational interactions can also connect them.
Within the class of 
renormalisable, gauge-invariant and mirror-symmetry-invariant operators, 
these interactions are
photon--mirror-photon kinetic mixing and Higgs-boson--mirror-Higgs-boson
coupling through a $\phi^{\dagger}\phi \phi'^{\dagger}\phi'$ term (we denote
mirror fields with a prime) \cite{parity2}. The implications of these interactions
have been explored in several papers \cite{parity2,epsilon}.

The mechanism under study here employs dimension-5 effective operators
of the form {\it lepton-Higgs-lepton$'$-Higgs$'$}
to couple the ordinary and mirror sectors. It also requires an assumption
about the initial lepton and baryon asymmetries (prior to reprocessing). 
In Ref.\ \cite{paper1}, a particular
such operator with natural initial asymmetries
(see below for a review) was found to yield
\begin{equation}
\frac{\Omega_B}{\Omega_{Dark}} \simeq 0.20 - 0.21,
\end{equation}
in excellent agreement with the allowed range suggested
by WMAP\cite{cmb} $0.20 \pm 0.02$.

We would now like to know if this encouraging quantitative success is unique
to the effective operator we happened to consider in Ref.\ \cite{paper1}. The purpose of
this paper is to examine a wider class of operators. We shall find that
while all selected operators yield $\Omega_B/\Omega_{Dark} < 1$, only a
subset are quantitatively consistent with the observations, and that the
operator of Ref.\ \cite{paper1} is not uniquely favoured (from
this cosmological perspective).

The next section reviews the dynamical framework of Ref.\ \cite{paper1} and
the section after that presents our calculations. We end by discussing
our results and drawing conclusions.

\section{Review of the dynamical framework}

A fundamental feature of mirror matter cosmology is that the temperature
of the mirror sector, $T'$, is expected to be 
different from that of the ordinary sector, $T$,
at least during certain epochs.
For the late epoch during which big bang nucleosynthesis takes place,
$T'/T \stackrel{<}{\sim} 0.5$ should hold in order to constrain the
expansion rate of the universe at that time. From a later period still,
large scale structure formation
with mirror DM suggests a slightly more stringent constraint, 
$T'/T \stackrel{<}{\sim} 0.2$ \cite{lss}. 
The smaller this ratio is, the more does
mirror DM resemble standard cold DM during the linear regime of
density perturbation growth (they obviously must differ in the non-linear
regime because mirror DM is self-interacting, chemically complex and 
dissipative) \cite{lss}.
For earlier epochs, however, we have no observational constraints that
$T'/T$ must satisfy. If $\Omega_{Dark} = \Omega'_B$, our fundamental hypothesis,
then the inequality of $\Omega'_B$ and $\Omega_B$ strongly suggests that
the temperatures were also different for at least part of the time 
during baryogenesis. 

A 
%initial 
temperature difference 
%(that is, at the big bang)
can be created, for example, through inflation \cite{inflation}. 
Imagine that there
is an inflaton and a mirror inflaton, and that inflation is
seeded by a fluctuation through whichever of these fields
the fluctuation ``favours''. Upon reheating, the {\it macroscopic}
ordinary-mirror asymmetry generated by the amplified fluctuation
will translate into $T' \neq T$ provided the two sectors are weakly
enough coupled to each other. The subsequent evolution of $T'/T$ depends
on the precise nature of the ordinary-mirror coupling terms. We
emphasise that the macroscopic temperature asymmetry is not at
all inconsistent with an exactly symmetric microphysics.

We can now review the main dynamic events our scenario requires: 
%(see Ref.\ \cite{paper1} for a more detailed introduction):

{\it Step 1.} Suppose that reheating after inflation leaves a
universe with 
\begin{equation}
T' \gg T,
\label{eq:TprimeggT}
\end{equation}
that is, the universe is totally dominated initially by mirror matter.
This would happen if inflation was driven by the vacuum
energy of a mirror inflaton.

{\it Step 2.} Around a certain temperature, $T' = T_1$, mirror baryon
and/or mirror lepton asymmetries are created. We prefer not to
specify the mechanism. It might be the out-of-equilibrium decays
of the heavy neutral mirror leptons as per leptogenesis \cite{leptogen}, or something
else. No significant ordinary-sector asymmetries are generated --
a reasonable assumption in the limit of Eq.(\ref{eq:TprimeggT}).

{\it Step 3.} These initial asymmetries are chemically reprocessed.
In particular, ordinary asymmetries are created through effective
dimension-5 operators of the form
\begin{equation}
L = \left( \frac{1}{M_N} \right)_{ij}\, \overline{\ell}_{iL}\,
\phi^c\, \ell'_{jR}\, \phi' + H.c.,
\label{eq:Lefffull}
\end{equation}
where $\ell_L$ is a left-handed ordinary lepton doublet, $\ell'_R$
is its mirror partner, $\phi$ is the ordinary Higgs doublet, and
$\phi'$ is its mirror partner. The indices $i,j=1,2,3$ denote 
the three families.
These are the lowest dimension non-renormalisable, gauge-invariant 
operators one can construct out of the fundamental fields. They may be 
generated from renormalisable operators through the exchange of
gauge-singlet fermions, hence the notation $M_N$. They also induce
terms  $m_{\nu} \equiv \langle\phi\rangle^2/M_N$ in the 
light neutrino mass
matrix. As well as contributing (with sphaleron and other
effects) to the chemical reprocessing of asymmetries, these interactions
also thermally equilibrate the sectors, that is, induce $T = T'$. 
They operate during the temperature regime,
\begin{equation}
M_N \stackrel{>}{\sim} T \stackrel{>}{\sim} T_2 \equiv 10^{10}(eV/m_{\nu})^2
\ GeV. 
\label{eq:LeffTrange}
\end{equation}
For $T \stackrel{>}{\sim} M_N$,
the effective Lagrangian approach is not valid and the parent fundamental,
renormalisable interactions are slower than the expansion rate, while
for $T \stackrel{<}{\sim} T_2$ the effective interactions are themselves 
slower than the expansion rate. The issue explored in this paper is how 
different family hierarchy assumptions for the 
$(1/M_N)_{ij}$ affect $\Omega_B/\Omega'_B$.

{\it Step 4.} A second but relatively brief inflationary episode (or some
alternative process) must then occur (beginning at some
temperature, $T_3$) in order to set up the mild
hierarchy, $T'/T \stackrel{<}{\sim}
0.2 - 0.5$, as required for big bang nucleosynthesis and the later
linear perturbation growth periods. It is interesting that this hierarchy
is in the opposite sense to that created by the first inflationary
episode [see Eq.\ (\ref{eq:TprimeggT})]. If the mirror inflaton induces
the earlier inflationary burst, then it is tempting to ascribe the
later burst to a largely failed attempt by the ordinary inflaton to reciprocate.

\section{Calculating $\Omega_B/\Omega_{Dark}$}

The main issue is the family structure assumed for Eq.\ (\ref{eq:Lefffull}).
We shall take for simplicity and definiteness that one flavour
combination for the effective operators dominates [through having
the largest $(1/M_N)_{ij}$], with all the others
too small to affect the chemical reprocessing. Given the connection with neutrino
mass, we shall assume the relevant $M_N$ is of order $\langle\phi\rangle^2/(1\ eV)$,
since the $\sim 1\ eV$ scale is a reasonable upper ``bound'' on neutrino mass,
and is even directly suggested by the LSND anomaly \cite{lsnd} [see also Ref.\ 
\cite{mirrornu} for discussions of ordinary-mirror neutrino mixing].

In Ref.\ \cite{paper1}, we examined only one case, where the $i = j = 2$
term dominates. 
Assuming that  there is just one eV mass term connecting
the ordinary and mirror sectors
and that it is approximately flavour diagonal, then
there are just 6 distinct cases to consider 
(including $i = j = 2$ to be called case 1 from now on):
\begin{eqnarray}
\text{case} & 1 &:\quad 
L = {1 \over M_N} \bar \ell_{2L} \phi^c \ell'_{2R} \phi' + H.c.
\label{eq:case1op}\\
\text{case} & 2 &:\quad 
L = {1 \over M_N} \bar \ell_{1L} \phi^c \ell'_{1R} \phi' + H.c.
\label{eq:case2op}\\
\text{case} & 3 &:\quad
L = {1 \over M_N} \bar \ell_{3L} \phi^c \ell'_{3R} \phi' + H.c.
\label{eq:case3op}\\
\text{case} & 4 &:\quad
L = {1 \over M_N} \bar \ell_{2L} \phi^c \ell'_{3R} \phi' + 
{1 \over M_N} \bar \ell'_{2R} \phi'^c \ell_{3L} \phi + H.c.
\label{eq:case4op}\\
\text{case} & 5 &:\quad
L = {1 \over M_N} \bar \ell_{1L} \phi^c \ell'_{3R} \phi' + 
{1 \over M_N} \bar \ell'_{1R} \phi'^c \ell_{3L} \phi + H.c.
\label{eq:case5op}\\
\text{case} & 6 &:\quad
L = {1 \over M_N} \bar \ell_{1L} \phi^c \ell'_{2R} \phi' + 
{1 \over M_N} \bar \ell'_{1R} \phi'^c \ell_{2L} \phi + H.c.
\label{eq:case6op}
\end{eqnarray}
These operators affect chemical reprocessing by constraining the chemical
potentials of the species concerned. We now need to review how the
reprocessing is analysed. 
%The $i = j = 1$ and $i = 1, j = 3$ cases give the
%same results as $i = j = 2$ 
%and $i = 2,  j = 3$, respectively (see next footnote).

For temperatures below about $10^{12}\ GeV$, QCD \cite{mohapatra}
and electroweak \cite{kuzmin} non-perturbative
processes
plus the Yukawa interactions for the fermions $c$, $\tau$, $b$ and $t$
are faster than the expansion rate of the universe \cite{buch}.
The Yukawa interactions
for $e_R$, $u_R$, $d_R$, $\mu_R$ and $s_R$ do not become fast enough until the
temperature drops below about $10^{10}\ GeV$.
%\footnote{Because both muon- and
%electron-Higgs Yukawa interactions are negligible above 
%$10^{10}\ GeV$, the statement
%in the last sentence of the previous paragraph holds.}  
Diagonal and off-diagonal weak interactions
involving left-handed quarks are also happening rapidly. By $T_2 \simeq 10^{10}\ GeV$
[see Eq.\ (\ref{eq:LeffTrange})], the selected effective operator, one of Eqs.\
(\ref{eq:case1op}-\ref{eq:case6op}), has also been inducing rapid interactions,
affecting the chemical composition and inducing $T = T'$. At about
$T = 10^{10}\ GeV$, all the quarks, leptons, Higgs bosons and their mirror
partners are in thermal equilibrium with distribution functions governed
by the temperature and chemical potentials for all the involved species.
We denote the chemical potential for species $X$ by $\mu_X$ ($X'$ by $\mu'_X$).
The rapid processes listed above relate the $\mu$'s. In addition,
we impose electric charge or hypercharge neutrality, and
mirror electric/hypercharge neutrality, for the universe.
 
The chemical constraint equations are \cite{buch}
\begin{eqnarray}
9\mu_q  + \sum_{i=1}^3 \mu_{\ell_i} &=& 0 \quad (\text{Electroweak  non-perturbative}),
\nonumber \\
6\mu_q  - \sum_{i=1}^3 (\mu_{u_i} + \mu_{d_i}) &=& 0 \quad (\text{QCD non-perturbative}),
\nonumber \\
3\mu_q  + 2\mu_{\phi} + \sum_{i=1}^3 
(2\mu_{u_i}-\mu_{d_i} - \mu_{\ell_i} - \mu_{e_i}) 
&=& 0 \quad (\text{Electric charge neutrality}),
\nonumber \\
\mu_q  - \mu_{\phi} - \mu_{d_3} &=& 0 \quad (\text{b-quark Yukawa}),
\nonumber \\
\mu_q  + \mu_{\phi} - \mu_{u_2} &=& 0 \quad (\text{c-quark Yukawa}),
\nonumber \\
\mu_q  + \mu_{\phi} - \mu_{u_3} &=& 0 \quad (\text{t-quark Yukawa}),
\nonumber \\
\mu_{\ell_3}  - \mu_{\phi} - \mu_{e_3} &=& 0 \quad (\text{$\tau$-lepton Yukawa}),
\label{eq:chemcons}
\end{eqnarray}
plus the corresponding seven equations from the mirror sector.
The simplified notation here is: $\mu_q \equiv \mu_{q_{1L}}
= \mu_{q_{2L}} = \mu_{q_{3L}}$, where the equalities are
enforced by off-diagonal weak interactions;
$\mu_{u_i} \equiv \mu_{u_{iR}}$, $\mu_{d_i} \equiv \mu_{d_{iR}}$
and $\mu_{\ell_i} \equiv \mu_{\ell_{iL}}$. We are working
in the Yukawa-diagonal basis (so $u_{3R}$ becomes the right-handed
component of the mass eigenstate $t$-quark after the electroweak
phase transition, and so on).

Then there are one or two more constraint equations, induced by the
dimension-5 operator(s):
\begin{eqnarray}
\text{case} & 1 &: - \mu_{\ell_2} - \mu_{\phi} + \mu_{\ell'_2} 
+ \mu_{\phi'} = 0;\label{eq:add1}\\
\text{case} & 2 &: - \mu_{\ell_1} - \mu_{\phi} + \mu_{\ell'_1} 
+ \mu_{\phi'} = 0;\label{eq:add2}\\
\text{case} & 3 &: - \mu_{\ell_3} - \mu_{\phi} + \mu_{\ell'_3} 
+ \mu_{\phi'} = 0;\label{eq:add3}\\
\text{case} & 4 &: - \mu_{\ell_2} - \mu_{\phi} + \mu_{\ell'_3} 
+ \mu_{\phi'} = 0;\ \ - \mu'_{\ell_2} - \mu'_{\phi} + \mu_{\ell_3} 
+ \mu_{\phi} = 0;\label{eq:add4}\\
\text{case} & 5 &: - \mu_{\ell_1} - \mu_{\phi} + \mu_{\ell'_3} 
+ \mu_{\phi'} = 0;\ \ - \mu'_{\ell_1} - \mu'_{\phi} + \mu_{\ell_3} 
+ \mu_{\phi} = 0;\label{eq:add5}\\
\text{case} & 6 &: - \mu_{\ell_1} - \mu_{\phi} + \mu_{\ell'_2} 
+ \mu_{\phi'} = 0;\ \ - \mu'_{\ell_1} - \mu'_{\phi} + \mu_{\ell_2} 
+ \mu_{\phi} = 0.\label{eq:add6}
\end{eqnarray}
So cases 1-3 have 15 constraint equations, while cases
4-6 have 16. With 28 chemical potential variables,
this leaves 13 free $\mu$'s for cases 1-3, and 12 for cases 4-6.
The free variables correspond to conserved quantities.

The solution strategy is simply to (i) select the appropriate number
of $\mu$'s (13 or 12) as independent variables, (ii) solve for the remaining
chemical potentials in terms of them, (iii) identify the independent, conserved
quantities for each case and write the associated chemical potential as a 
linear combination of the independent chemical potentials chosen at step (i).
Finally, (iv) solve for the baryon/lepton and mirror baryon/lepton 
asymmetries at $T \simeq 10^{10}\ GeV$
in terms of the conserved charges. Below $10^{10}\ GeV$, other
processes come into play in determining the final, low temperature 
values of the baryon and mirror baryon asymmetries via a by now 
standard procedure (see below).

We now examine each case defined by 
Eqs.\ (\ref{eq:case1op}-\ref{eq:case6op}) separately. 
In the main text, we shall write down
the conserved charges and give the expressions for $B$, $L$, $B'$ and $L'$
at $T \simeq 10^{10}\ GeV$ in terms of those charges. In the Appendix, we give
the algebraic details for case 1 to illustrate the methodology.

\subsection{Cases 1 \& 2}

Considering first case 1, at $T \sim 10^{10}$ GeV there are
13 conserved charges:
\begin{eqnarray}
&{\cal L}_0 = {1 \over 3} B - L_2 + {1 \over 3} B' - L_2' &.
\nonumber \\
& {\cal L}_1 = {1 \over 3} B - L_1,\ \
{\cal L}_1' = {1 \over 3} B' - L_1',&
\nonumber \\
& {\cal L}_2 = {1 \over 3}B - L_3,\ \
{\cal L}_2' = {1 \over 3}B' - L_3',&
\nonumber \\
&{\cal L}_3 = L_{e_{1R}}, \ \ \ \
{\cal L}_3' = L'_{e_{1R}},&
\nonumber \\
&{\cal L}_4 =  L_{e_{2R}}, \ \ \ \
{\cal L}_4' = L'_{e_{2R}},&
\nonumber \\
&{\cal L}_5 = B_{u_{1R}} - B_{d_{1R}}, \ \
{\cal L}_5' = B'_{u_{1R}} - B'_{d_{1R}},&
\nonumber \\
&{\cal L}_6 = B_{d_{1R}} - B_{d_{2R}},\ \
{\cal L}_6' = B'_{d_{1R}} - B'_{d_{2R}}.&
\label{eq:case1conserved}
\end{eqnarray}
The notation is: 
$L_i$ is 
the lepton number of family $i$, $B$
($L = L_1 + L_2 + L_3$) is total baryon(lepton) number, 
$B_{q_{iR}}$ is
the charge for right-handed quark $q$ from family $i$, and $L_{e_{iR}}$
is the charge for the right-handed charged lepton of family $i$. The
respective quantities from the mirror sector carry a prime.

In terms of these quantities,
\begin{eqnarray}
B &=&  \alpha_0 {\cal L}_0 +
\sum_{i=1}^6 \alpha_i {\cal L}_i + \sum_{i=1}^6 \alpha'_i {\cal L'}_i,
\nonumber \\
L &=& \beta_0 {\cal L}_0 + \sum_{i=1}^6 \beta_i {\cal L}_i + 
\sum_{i=1}^6 \beta'_i {\cal L'}_i .
\label{eq:BandLcase1}
\end{eqnarray}
Under mirror symmetry, $B \leftrightarrow B'$, $L
\leftrightarrow L'$, ${\cal L}_i \leftrightarrow {\cal L}'_i$
(and ${\cal L}_0 \to {\cal L}_0$). Hence,
\begin{eqnarray}
B' &=& \alpha_0 {\cal L}_0 + \sum_{i=1}^6 \alpha_i {\cal L'}_i 
+\sum_{i=1}^6 \alpha_i' {\cal L}_i,
\nonumber \\ 
L' &=& \beta_0 {\cal L}_0 + \sum_{i=1}^6 \beta_i {\cal L'}_i 
+ \sum_{i=1}^6 \beta_i' {\cal L}_i.
\label{eq:BandLprimecase1}
\end{eqnarray}
The values of the $\alpha$ and $\beta$ parameters
are given in Table \ref{table:case1}.

Results for case 2 follow from
case 1 with the replacements:
$L_2 \to L_1, \ L'_2 \to L'_1$ in ${\cal L}_0$ and
$L_1$ ($L'_1$) $\to$ $L_2$ ($L'_2$) in ${\cal L}_1$ (${\cal L}'_1$).
Because both muon- and
electron-Higgs Yukawa interactions are negligible above 
$10^{10}\ GeV$, 
the $\alpha, \beta$ coefficients for case 2 are trivially the same 
as case 1 which were given in Table \ref{table:case1}.

%%%%%%%%%%%%%%%%%%%%%%%%%%% Table %%%%%%%%%%%%%%%%%%%%%%%%%%%%%%%%%

%\vskip 1cm
%\begin{center}
%The $\alpha$ and $\beta$ coefficients
%in the expansions of $B$ and $L$ defined in Eqs.\ (\ref{eq:BandLcase1}) and 
%(\ref{eq:BandLprimecase1}).
%\end{center}
%\begin{center}
\begin{table*}
\caption{\label{table:case1}The $\alpha$ and $\beta$ coefficients
in the expansions of $B$ and $L$ for cases 1 \& 2 
as defined in Eqs.\ (\ref{eq:BandLcase1}) and 
(\ref{eq:BandLprimecase1}).}
\begin{ruledtabular}
\begin{tabular}{|c|c|c|c|}
%\begin{tabular*}
%{0.75\textwidth}{@{\extracolsep{\fill}}
%|c|c|c|c|}  \hline
%%\setlength{\tabcolsep}{1pt} 
& & & \\
$\alpha_0 = \frac{69}{316}$ &   &
$\beta_0 = \frac{-89}{316}$ &  \\
& & &
\\
$\alpha_1 = \frac{43263}{98276}$ & $\alpha'_1 = \frac{-345}{98276}$ &
$\beta_1 = \frac{-55803}{98276}$ & $\beta'_1 = \frac{445}{98276}$ \\
& & &
\\
$\alpha_2 = \frac{7050}{24569}$ & $\alpha'_2 = \frac{414}{24569}$ &
$\beta_2 = \frac{-16571}{24569}$ & $\beta'_2 = \frac{-534}{24569}$ \\
& & &
\\
$\alpha_3 = \frac{45189}{98276}$ & $\alpha'_3 = \frac{-6003}{98276}$ &
$\beta_3 = \frac{31443}{98276}$ & $\beta'_3 = \frac{7743}{98276}$ \\
& & &
\\
$\alpha_4 = \frac{23385}{98276}$ & $\alpha'_4 = \frac{15801}{98276}$ &
$\beta_4 = \frac{59567}{98276}$ & $\beta'_4 = \frac{-20381}{98276}$ \\
& & &
\\
$\alpha_5 = \frac{-963}{24569}$ & $\alpha'_5 = \frac{2829}{24569}$ &
$\beta_5 = \frac{5515}{24569}$ & $\beta'_5 = \frac{-3649}{24569}$ \\
& & &
\\
$\alpha_6 = \frac{-963}{49138}$ & $\alpha'_6 = \frac{2829}{49138}$ &
$\beta_6 = \frac{5515}{49138}$ & $\beta'_6 = \frac{-3649}{49138}$ \\
& & & \\
%\hline
%\label{table:case1}
%\caption{The $\alpha$ and $\beta$ coefficients
%in the expansions of $B$ and $L$ defined in Eqs.\ (\ref{eq:BandLcase1}) and 
%(\ref{eq:BandLprimecase1}).}
%\end{tabular*}
%\end{center}
%\vskip 0.8cm
\end{tabular}
\end{ruledtabular}
\end{table*}
%%%%%%%%%%%%%%%%%%%%%%%%%%%%%%%%%%%%%%%%%%%%%%%%%%%%%%%%%%%%5

\subsection{Case 3}

In this case, we again have 13 conserved charges,
as in Eq.\ (\ref{eq:case1conserved}), but with
$L_3 \to L_2$ 
($L_3' \to L_2'$) 
in ${\cal L}_2$ (${\cal L}'_2$), and $L_2 \to L_3,\
L_2' \to L_3'$ in ${\cal L}_0$. The $\alpha$ and $\beta$
values are listed in Table \ref{table:case2}.

%% table 2 here %%
\begin{table*}
\caption{\label{table:case2}The $\alpha$ and $\beta$ coefficients
in the expansions of $B$ and $L$ for case 3.}
\begin{ruledtabular}
\begin{tabular}{|c|c|c|c|}
%\begin{center}
%\begin{tabular*}
%{0.75\textwidth}{@{\extracolsep{\fill}}
%|c|c|c|c|}  \hline
%%\setlength{\tabcolsep}{1pt} 
& & & \\
$\alpha_0 = \frac{12}{79}$ &   &
$\beta_0 = \frac{-55}{158}$ &  \\
& & &
\\
$\alpha_1 = \alpha_2 = \frac{735}{1738}$ & $\alpha'_1 = \alpha'_2 = 
\frac{12}{869}$ &
$\beta_1 = \beta_2 = \frac{-42}{79}$ & $\beta'_1 = \beta'_2 = 
\frac{-5}{158}$ \\
& & &
\\
$\alpha_3 = \alpha_4 = \frac{71}{158}$ & $\alpha'_3 = \alpha'_4 = 
\frac{-4}{79}$ &
$\beta_3 = \beta_4 = \frac{67}{237}$ & $\beta'_3 = \beta'_4 =
\frac{55}{474}$ \\
& & &
\\
$\alpha_5 = \frac{-46}{869}$ & $\alpha'_5 = \frac{112}{869}$ &
$\beta_5 = \frac{88}{237}$ & $\beta'_5 = \frac{-70}{237}$ \\
& & &
\\
$\alpha_6 = \alpha_5/2$ & $\alpha'_6 = \alpha'_5/2$ &
$\beta_6 = \beta_5/2$ & $\beta'_6= \beta'_5/2$ \\
& & &
\\
\end{tabular}
\end{ruledtabular}
\end{table*}

%\hline
%\label{table:case2}
%\end{tabular*}
%\end{center}
%\vskip 0.8cm
%%\noindent

\subsection{Cases 4 \& 5}

Considering first case 4, at $T \sim 10^{10}$ GeV,
there are 12 conserved charges:
\begin{eqnarray}
&{\cal L}_0 = \frac{1}{3} B - L_3 + {1 \over 3} B' - L_2',\ \
{\cal L}_0' = \frac{1}{3} B' - L_3' + {1 \over 3} B - L_2,&
\nonumber \\
&{\cal L}_1 = {1 \over 3} B - L_1,\ \
{\cal L}_1' = {1 \over 3} B' - L_1',&
\nonumber \\
&{\cal L}_2 = L_{e_{1R}}, \ \ \ \
{\cal L}_2' = L'_{e_{1R}},&
\nonumber \\
&{\cal L}_3 =  L_{e_{2R}}, \ \ \ \
{\cal L}_3' = L'_{e_{2R}},&
\nonumber \\
&{\cal L}_4 = B_{u_{1R}} - B_{d_{1R}}, \ \
{\cal L}_4' = B'_{u_{1R}} - B'_{d_{1R}},&
\nonumber \\
&{\cal L}_5 = B_{d_{1R}} - B_{d_{2R}},\ \
{\cal L}_5' = B'_{d_{1R}} - B'_{d_{2R}}.&
\label{eq:case3conserved}
\end{eqnarray}
In terms of these quantities,
\begin{eqnarray}
B &=&
\sum_{i=0}^5 \left(\alpha_i {\cal L}_i + \alpha'_i {\cal L'}_i \right),
\nonumber \\
L &=& \sum_{i=0}^5 \left(\beta_i {\cal L}_i + \beta'_i {\cal L'}_i \right).
\label{eq:BandLcase3}
\end{eqnarray}
Under mirror symmetry, $B \leftrightarrow B'$, $L
\leftrightarrow L'$, ${\cal L}_i \leftrightarrow {\cal L}'_i$. 
Hence,
\begin{eqnarray}
B' &=& \sum_{i=0}^5 \left(\alpha_i {\cal L'}_i 
+ \alpha_i' {\cal L}_i\right),
\nonumber \\ 
L' &=& \sum_{i=0}^5 \left(\beta_i {\cal L'}_i 
+ \beta_i' {\cal L}_i\right).
\label{eq:BandLprimecase3}
\end{eqnarray}
The values of the $\alpha$ and $\beta$ parameters
are given in Table \ref{table:case3}.

%%%%%%%%%%%%%%%%%%%%%%%%%%% Table %%%%%%%%%%%%%%%%%%%%%%%%%%%%%%%%%
\begin{table*}
\caption{\label{table:case3}The $\alpha$ and $\beta$ coefficients
in the expansions of $B$ and $L$ for cases 4 \& 5 as defined in 
Eqs.\ (\ref{eq:BandLcase3}) and 
(\ref{eq:BandLprimecase3}).}
\begin{ruledtabular}
\begin{tabular}{|c|c|c|c|}
%\vskip 1cm
%\begin{center}
%\begin{tabular*}
%{0.75\textwidth}{@{\extracolsep{\fill}}
%|c|c|c|c|}  \hline
%\setlength{\tabcolsep}{1pt} 
& & & \\
D = 328321 & & & 
\\
$\alpha_0 = \frac{53484}{D}$ &   
$\alpha'_0 = \frac{66132}{D}$ &
$\beta_0 = \frac{-123061}{D}$ 
& $\beta'_0 = \frac{-85644}{D} $
\\
& & &
\\
$\alpha_1 = \frac{140874}{D}$ & $\alpha'_1 = \frac{4908}{D}$ &
$\beta_1 = \frac{-177434}{D}$ & $\beta'_1 = \frac{-5105}{D}$ \\
& & &
\\
$\alpha_2 = \frac{165018}{D}$ & $\alpha'_2 = \frac{-34188}{D}$ &
$\beta_2 = \frac{72793}{D}$ & $\beta'_2 = \frac{58037}{D}$ \\
& & &
\\
$\alpha_3 = \frac{90276}{D}$ & $\alpha'_3 = \frac{14388}{D}$ &
$\beta_3 = \frac{164583}{D}$ & $\beta'_3 = \frac{-59919}{D}$ \\
& & &
\\
$\alpha_4 = \frac{-48288}{D}$ & $\alpha'_4 = \frac{78192}{D}$ &
$\beta_4 = \frac{156188}{D}$ & $\beta'_4 = \frac{-126284}{D}$ \\
& & &
\\
$\alpha_5 = \frac{\alpha_4}{2}$ & $\alpha'_5 = \frac{\alpha'_4}{2}$ &
$\beta_5 = \frac{\beta_4}{2}$ & $\beta'_5 = \frac{\beta'_4}{2}$ \\
& & &
\\
\end{tabular}
\end{ruledtabular}
\end{table*}
%\hline
%\label{table:case3}
%\end{tabular*}
%\end{center}
%\vskip 0.8cm
%%\noindent
%%%%%%%%%%%%%%%%%%%%%%%%%%%%%%%%%%%%%%%%%%%%%%%%%%%%%%%%%%%%5

Results for case 5 follow from
case 4 with the replacements:
$L'_2 \to L'_1$ ($L_2 \to L_1$) in ${\cal L}_0$ (${\cal L}'_0$)
and $L_1 \to L_2$ ($L'_1 \to L'_2$) in ${\cal L}_1$ (${\cal L}'_1$).
Because both muon- and
electron-Higgs Yukawa interactions are negligible above 
$10^{10}\ GeV$, the
$\alpha, \beta$ coefficients for case 5 are trivially the same 
as those of case 4 (given in Table \ref{table:case3}).

\subsection{Case 6}

This case is similar to case 4, except that 
$L_1 \to L_3$ ($L_1' \to L_3'$) in ${\cal L}_1$
(${\cal L}'_1$) and
$L_3 \to L_1$ ($L_3' \to L_1'$) in ${\cal L}_0$
(${\cal L}'_0$).
The $\alpha, \beta$ coefficients are given in Table \ref{table:case4}

%%%%%%%%%%%%%%%%%%%%%%%%%%% Table %%%%%%%%%%%%%%%%%%%%%%%%%%%%%%%%%

\begin{table*}
\caption{\label{table:case4}The $\alpha$ and $\beta$ coefficients
in the expansions of $B$ and $L$ for case 6.}
\begin{ruledtabular}
\begin{tabular}{|c|c|c|c|}
%\vskip 1cm
%\begin{center}
%\begin{tabular*}
%{0.75\textwidth}{@{\extracolsep{\fill}}
%|c|c|c|c|}  \hline
%%\setlength{\tabcolsep}{1pt} 
& & & \\
$D = 1343$ 
& $\alpha_0 = \alpha'_0 = \frac{3519}{12D}$ &   
$D_2 \equiv 3D = 4029$
&
$\beta_0 = \beta'_0 = \frac{-13617}{12D_2}$   \\
& & &
\\
$\alpha_1 = \frac{362}{D}$ & $\alpha'_1 = \frac{46}{D}$ &
$\beta_1 = \frac{-2627}{D_2}$ & $\beta'_1 = \frac{-178}{D_2}$ \\
& & &
\\
$\alpha_2 = \alpha_3 = \frac{4793}{12D}$ & $\alpha'_2 =
\alpha'_3 = \frac{1633}{12D}$ &
$\beta_2 = \beta_3 = \frac{25597}{12D_2}$ & $\beta'_2 = \beta'_3 =
\frac{-6319}{12D_2}$ \\
& & &
\\
$\alpha_4 = \frac{-637}{3D}$ & $\alpha'_4 = \frac{943}{3D}$ &
$\beta_4 = \frac{4567}{3D_2}$ & $\beta'_4 = \frac{-3649}{3D_2}$ \\
& & &
\\
$\alpha_5 = \frac{\alpha_4}{2}$ & $\alpha'_5 = \frac{\alpha'_4}{2}$ &
$\beta_5 = \frac{\beta_4}{2}$ & $\beta'_5 = \frac{\beta'_4}{2}$ \\
& & &
\\
\end{tabular}
\end{ruledtabular}
\end{table*}
%\hline
%\label{table:case4}
%\end{tabular*}
%\end{center}
%\vskip 0.8cm
%%\noindent
%%%%%%%%%%%%%%%%%%%%%%%%%%%%%%%%%%%%%%%%%%%%%%%%%%%%%%%%%%%%5

\section{Results and conclusions}

The values of the conserved charges,
${\cal L}_i$, ${\cal L}'_i$ and ${\cal L}_0$
depend on the {\it initial} asymmetry generation mechanism.
As in Ref.\ \cite{paper1}, we consider, for definiteness, the simple case of
non-zero $B'$ and/or $L'$:
$B' = X_0'$, $L' = Y_0'$, $B = L = 0$ (with $L'_{\ell_{1}} =
L'_{\ell_{2}} = L'_{\ell_{3}} \equiv Y_0'/3$ and $B'_{u_{1R}} = B'_{d_{1R}} =
B'_{d_{2R}}$).
The only nonzero conserved charges are then
\begin{eqnarray}
{\cal L}'_1 &=& {\cal L}'_2 = {\cal L}_0 = \frac{1}{3}(X_0' - Y_0')
\equiv Z \ \ \  {\rm cases \ 1-3},
\nonumber \\
{\cal L}'_1 &=& {\cal L}'_0 = {\cal L}_0 \equiv Z \ \ \ 
{\rm cases \ 4-6}.
\end{eqnarray}

After chemical processing, the baryon and mirror baryon asymmetries
at $T \simeq 10^{10}\ GeV$ are then
\begin{eqnarray} B &=& Z (\alpha_0 + \alpha'_1  + \alpha'_2), 
\ B' = Z (\alpha_0 + \alpha_1 + \alpha_2),
\ \  {\rm cases\ 1-3}
\nonumber \\
B &=& Z (\alpha_0 + \alpha'_0 + \alpha'_1), \ B' = Z (\alpha_0 + 
\alpha'_0 + \alpha_1),
\ \ \  {\rm cases\  4-6}.
\end{eqnarray}
The ordinary matter/dark matter ratio at that temperature is therefore
\begin{eqnarray}
{B \over B'} &=& {\alpha_0 + \alpha'_1  + \alpha'_2 \over 
\alpha_0 + \alpha_1 + \alpha_2} 
= {3795 \over 15487}
\simeq 0.25
\ \ \ {\rm cases\ 1 \ \& \ 2},
\nonumber \\
{B \over B'} &=& {\alpha_0 + \alpha'_1  + \alpha'_2 \over 
\alpha_0 + \alpha_1 + \alpha_2} 
= {52 \over 289}
\simeq 0.18
\ \ \ {\rm case\ 3},
\nonumber \\
{B \over B'} &=& {\alpha_0 + \alpha'_0 + \alpha'_1  \over \alpha_0 +
\alpha'_0 + \alpha_1 } 
= {20754 \over 43415}
\simeq 0.48
\ \ \ {\rm cases\ 4 \ \& \ 5},
\nonumber \\
{B \over B'} &=& {\alpha_0 + \alpha'_0 + \alpha'_1  \over \alpha_0 +
\alpha'_0 + \alpha_1 } 
= {1265 \over 1897}
\simeq 0.67
\ \ \ {\rm case\ 6}.
\label{22}
\end{eqnarray}

The value of this ratio changes to some 
extent at lower temperatures as different
chemical processes become important.
However, at temperatures near that of
the electroweak phase transition, $T = T_{EW} \sim 200$ GeV,
the values of $B$ and $B'$ depend only on
the values of $B-L$ and $B'-L'$.
These charges are separately conserved for $T \ll 10^{10}$ GeV,
because the interactions in Eqs.\ (\ref{eq:case1op}-\ref{eq:case4op}) chemically connecting
the ordinary and mirror sectors are slower than the expansion rate.
This yields the well-known relation between $B$ and $B-L$ \cite{shap},
\begin{eqnarray}
B = {28 \over 79} (B - L),
\end{eqnarray}
with an identical relation for $B'$ in terms of $B'-L'$.
Below the electroweak phase transition temperature, there are no processes
fast enough to further affect $B$ and $B'$.
Thus the final, low temperature value of the ratio $B/B' = \Omega_B/\Omega'_B$ is
simply given $(B-L)/(B'-L')$ evaluated at $T \simeq 10^{10}\ GeV$:
\begin{eqnarray}
{\Omega_B \over \Omega'_B} &=& {\alpha_0 + \alpha'_1 + \alpha'_2 - \beta_0 - \beta'_1
- \beta'_2
\over \alpha_0 + \alpha_1 + \alpha_2 - \beta_0 - \beta_1 - \beta_2}
= {55 \over 256}
\simeq 0.22,\ \ \ {\rm cases\ 1 \ \& \ 2}; \nonumber \\
{\Omega_B \over \Omega'_B} &=& {\alpha_0 + \alpha'_1 + \alpha'_2 - \beta_0 - \beta'_1
- \beta'_2
\over \alpha_0 + \alpha_1 + \alpha_2 - \beta_0 - \beta_1 - \beta_2}
= {13 \over 53}
\simeq 0.25, \ \ \ {\rm case\ 3}; \nonumber \\
{\Omega_B \over \Omega'_B} &=& {\alpha_0 + \alpha'_0 + \alpha'_1  - \beta_0 -
\beta'_0 - \beta'_1
\over \alpha_0 + \alpha'_0 + \alpha_1  - \beta_0 - \beta'_0 - \beta_1 }
= {214 \over 409}
\simeq 0.52, 
\ \ \ {\rm cases\ 4 \ \& \ 5};
\nonumber \\
{\Omega_B \over \Omega'_B} &=& {\alpha_0 + \alpha'_0 + \alpha'_1  - \beta_0 -
\beta'_0 - \beta'_1
\over \alpha_0 + \alpha'_0 + \alpha_1  - \beta_0 - \beta'_0 - \beta_1 }
=
{55 \over 98}
\simeq 0.56,
\ \ \ {\rm case\ 6}.
\label{eq:finalresults1}
\end{eqnarray}

If there is some brief period of inflation between
$T \sim 10^{10}\ GeV$ and $T_{EW}$, as suggested by Step 4, 
then the results depend on when this
second period of inflation occurs, $T=T_3$, as well as
the subsequent ordinary and mirror sector reheating temperatures, $T_{RH}$ and
$T'_{RH}$ respectively (with $T'_{RH} < T_{RH}$ required for successful
big bang nucleosynthesis and large scale structure formation). 

At one extreme, if $T_3 \sim T_{EW}$, then the results of Eq.\ (\ref{eq:finalresults1})
hold irrespective of the reheating temperatures. The other extreme case, $T_3 = 10^{10}\ GeV$,
allows three outcomes. If $T_{RH}>T_{EW}$ and $T'_{RH}>T_{EW}$, then Eq.\ (\ref{eq:finalresults1})
again holds. In the opposite situation, $T_{RH}<T_{EW}$ and $T'_{RH}<T_{EW}$, the final
values are simply given by Eq.\ (\ref{22}), because no further reprocessing can take place.
The acceptable intermediate situation, $T_{RH}>T_{EW}$ and $T'_{RH}<T_{EW}$, has further
ordinary sector reprocessing but a frozen mirror sector. For this situation, the
final ratios are given by
\begin{eqnarray}
{\Omega_B \over \Omega'_B} &=& {28 \over 79}{\alpha_0 + \alpha'_1 + \alpha'_2
-\beta_0 - \beta'_1 - \beta'_2 \over \alpha_0 + \alpha_1 + \alpha_2}
= {3080 \over 15487}
\simeq 0.20, \ \ {\rm cases \ 1 \ \& \ 2}; \nonumber \\
{\Omega_B \over \Omega'_B} &=& {28 \over 79}{\alpha_0 + \alpha'_1 + \alpha'_2
-\beta_0 - \beta'_1 - \beta'_2 \over \alpha_0 + \alpha_1 + \alpha_2}
= {182 \over 867}
\simeq 0.21, \ \ {\rm case \ 3}; \nonumber \\
{\Omega_B \over \Omega'_B} &=& {28 \over 79}{\alpha_0 + \alpha'_0 + \alpha'_1
-\beta_0 - \beta'_0 - \beta'_1 \over \alpha_0 + \alpha'_0 + \alpha_1}
= {1578892 \over 3429785}
\simeq 0.46, \ \ {\rm cases \ 4 \ \& \ 5}; \nonumber \\
{\Omega_B \over \Omega'_B} &=& {28 \over 79}{\alpha_0 + \alpha'_0 + \alpha'_1
-\beta_0 - \beta'_0 - \beta'_1 \over \alpha_0 + \alpha'_0 + \alpha_1}
= {440 \over 813}
\simeq 0.54, \ \ {\rm case \ 6}.
\label{eq:finalresults2} 
\end{eqnarray}

These results all reproduce the qualitative observation that there is more dark
matter than ordinary matter. Intuitively this is because
the FRW universe is born full of mirror matter (under
our assumptions), and only some of the net mirror baryon/lepton number
is chemically reprocessed into an ordinary baryon asymmetry. But it is also
interesting that there is a subset of effective dimension-5 operators
which are
{\it quantitatively} successful, namely cases 1-3 [with case 3 
marginal unless
the circumstances leading to Eq.\ (\ref{eq:finalresults2}) obtain]. Given that
the effective operators also contribute to the light neutrino mass matrix
through ordinary-mirror neutrino mixing, a tentative connection between
the dark matter problem and neutrino oscillation physics can be made. The
connection must be tentative, because some important assumptions lie behind
our results, especially the microphysical desert between the electroweak
scale and physics at $10^{10-12}\ GeV$ (as emphasised in Ref.\ \cite{paper1}).

In conclusion, we have shown that the ratio of baryonic to non-baryonic
dark matter, $\Omega_{b}/\Omega_{dark} = 0.20 \pm 0.02$, inferred by 
WMAP\cite{cmb}, can be {\it quantitatively} explained if mirror
matter is identified with the non-baryonic dark matter.
Our explanation involves a set of assumptions about the physics
governing the early evolution of the Universe, which are not
unique but are nevertheless plausible. Our approach also has 
important implications for neutrino physics, suggesting eV scale
neutrino masses, which can be tested/constrained from upcoming neutrino
experiments such as miniBooNE.

\acknowledgments{This work was supported by the Australian Research Council.}

\appendix

\section{Solving the case 1 equations}

We show the algebraic technique for solving case 1. The other cases
follow similarly. 

The aim is to use Eqs.\ (\ref{eq:chemcons}) and (\ref{eq:add1})
to extract the total baryon number $B$, total lepton number $L$, and their
mirror matter analogues $B'$ and $L'$ at $T \simeq 10^{10}\ GeV$.
These 15 equations reduce the 28 variables, $\mu_q$, $\mu_{u_i}$, 
$\mu_{d_i}$, $\mu_{\ell_i}$, $\mu_{e_i}$, $\mu_{\phi}$ and their
primed counterparts to 13 independent variables. The number of independent
variables corresponds to the number of conserved charges,
Eq.\ (\ref{eq:case1conserved}).
The problem at hand is to find $B$, $L$, $B'$ and $L'$ in terms
of the conserved charges.

One systematic way of doing this is the following.
First identify 13 independent variables. One possible
choice is the following: $\mu_q$, $\mu_{\ell_2}$, $\mu_{\ell_3}$,
$\mu_{e_1}$, $\mu_{e_2}$, $\mu_{d_1}$, $\mu_{d_2}$, $\mu'_{q}$, $\mu'_{\ell_3}$,
$\mu'_{e_1}$, $\mu'_{e_2}$, $\mu'_{d_1}$ and $\mu'_{d_2}$. 
Then use Eqs.\ (\ref{eq:chemcons}) and (\ref{eq:add1}) to
write the 15 dependent $\mu$ variables in terms of the
chosen independent variables. Doing this we have:
\begin{eqnarray}
\mu_{\ell_1} &=& -9\mu_q - \mu_{\ell_2} - \mu_{\ell_3}
\nonumber \\
\mu_{\phi} &=& {1 \over 6} \left( -21\mu_q + 3\mu_{d_1} + 3\mu_{d_2} + \mu_{e_1}
+ \mu_{e_2} + \mu_{\ell_3} \right)
\nonumber \\
\mu_{d_3} &=& {1 \over 6} \left( 27 \mu_q - 3\mu_{d_1} - 3\mu_{d_2} - \mu_{e_1}
- \mu_{e_2} - \mu_{\ell_3} \right)
\nonumber \\
\mu_{e_3} &=& {1 \over 6} \left( 21\mu_q - 3\mu_{d_1} - 3\mu_{d_2} - \mu_{e_1} -
\mu_{e_2} + 5\mu_{\ell_3} \right)
\nonumber \\
\mu_{u_3} &=& \mu_{u_2} = {1 \over 6} \left( -15\mu_q + 3\mu_{d_1} + 3\mu_{d_2}
+ \mu_{e_1} + \mu_{e_2} + \mu_{\ell_3} \right)
\nonumber \\
\mu_{u_1} &=& {1 \over 6} \left( 39\mu_q - 9\mu_{d_1} - 9\mu_{d_2} - \mu_{e_1} -
\mu_{e_2} - \mu_{\ell_3} \right)
\nonumber \\
\mu'_{\ell_2} &=&  \mu_{\ell_2} + \mu_{\phi} -\mu'_{\phi}
\nonumber \\
\mu'_{\ell_1} &=& -9\mu'_{q} - \mu_{\ell_2} - \mu_{\phi} + \mu'_{\phi}
-\mu'_{\ell_3}
\nonumber \\
\mu'_{\phi} &=& {1 \over 6} \left( -21\mu'_{q} + 3\mu'_{d_1} + 3\mu'_{d_2} +
\mu'_{e_1} + \mu'_{e_2} + \mu'_{\ell_3} \right)
\nonumber \\
\mu'_{d_3} &=& {1 \over 6} \left( 27 \mu'_{q} - 3\mu'_{d_1} - 3\mu'_{d_2} -
\mu'_{e_1} - \mu'_{e_2} - \mu'_{\ell_3} \right)
\nonumber \\
\mu'_{e_3} &=& {1 \over 6} \left( 21\mu'_{q} - 3\mu'_{d_1} - 3\mu'_{d_2} - 
\mu'_{e_1} - \mu'_{e_2} + 5\mu'_{\ell_3} \right)
\nonumber \\
\mu'_{u_3} &=& \mu'_{u_2} = {1 \over 6} \left( -15\mu'_{q} + 3\mu'_{d_1} +
3\mu'_{d_2} + \mu'_{e_1} + \mu'_{e_2} + \mu'_{\ell_3} \right)
\nonumber \\
\mu'_{u_1} &=& {1 \over 6} \left( 39\mu'_{q} - 9\mu'_{d_1} - 9\mu'_{d_2} -
\mu'_{e_1} - \mu'_{e_2} - \mu'_{\ell_3} \right),
\label{eq:depindep}
\end{eqnarray}
where it is understood that in the $\mu'_{\ell_{1,2}}$ equations, $\mu_{\phi}$
and $\mu'_{\phi}$ are to be substituted with the respective righthand
side's above. In terms of the chosen independent $\mu_i$, the
baryon and lepton number are:
\begin{eqnarray}
B &=& 6\mu_q + \sum_{i=1}^3 \left(\mu_{u_i} + \mu_{d_i}\right)
\nonumber \\
&=& 12\mu_q
\nonumber \\
L &=& \sum_{i=1}^3 \left( 2\mu_{\ell_i} + \mu_{e_i}
\right)
\nonumber \\
&=& {-29 \over 2}\mu_q - {1 \over 2}\mu_{d_1} - {1\over 2}\mu_{d_2}
+ {5 \over 6}\mu_{e_1} + {5 \over 6}\mu_{e_2} + {5 \over 6}\mu_{\ell_3}
\label{lb}
\end{eqnarray}

The next step in the calculation is to write the conserved
charges, ${\cal L}_i$ in Eq.\ (\ref{eq:case1conserved}), 
in terms of the 13 independent variables.
For example:\footnote{The conventional definition of, say,
the ``baryon number of the universe'' is the ratio $n_B/s$
where $n_B$ is the net baryon number per unit volume, while
$s$ is entropy density. There is proportionality factor
relating this definition of baryon number to the simple
one convenient for our application (just the appropriate linear
combination of chemical potentials).} 
\begin{eqnarray}
{\cal L}_1 &=& {1 \over 3}B - L_1
\nonumber \\
&=& 
2\mu_q + {1 \over 3} \sum_{i=1}^3 (\mu_{u_i} + \mu_{d_i})
- 2\mu_{\ell_1} - \mu_{e_1}
\nonumber \\
&=& 22\mu_q + 2\mu_{\ell_2} + 2\mu_{\ell_3} - \mu_{e_1},
\end{eqnarray}
where Eq.\ (\ref{eq:depindep}) has been used in the last step.
The results for the other ${\cal L}_i$ are given below:
\begin{eqnarray}
{\cal L}_0 &=&
11 \mu_q - 4\mu_{\ell_2} - {4 \over 3} \mu_{e_2} - {1 \over 3} \mu_{e_1}
- {1 \over 3}\mu_{\ell_3} - \mu_{d_1} - \mu_{d_2}
- 3\mu'_q + \mu'_{d_1} + \mu'_{d_2} + {1 \over 3} \mu'_{e_1}
- {2 \over 3} \mu'_{e_2} + {1 \over 3} \mu'_{\ell_3} 
\nonumber \\
{\cal L}_2 &=& {1 \over 6}\left( 3\mu_q + 3\mu_{d_1} + 3\mu_{d_2}
+ \mu_{e_1} + \mu_{e_2} - 17\mu_{\ell_3} \right)
\nonumber \\
{\cal L}_3 &=&
\mu_{e_1} 
\nonumber \\
{\cal L}_4 &=&
\mu_{e_2} 
\nonumber \\
{\cal L}_5 &=&
{1 \over 6}\left(
39\mu_q - 15\mu_{d_1} - 9\mu_{d_2} - \mu_{e_1} - \mu_{e_2} -
\mu_{\ell_3}
\right)
\nonumber \\
{\cal L}_6 &=&
\mu_{d_1} - \mu_{d_2}
\nonumber \\
{\cal L}'_1 &=&
29\mu'_q - 7\mu_q + \mu_{d_1} + \mu_{d_2} + {1 \over 3}\mu_{e_1}
+ {1\over 3}\mu_{e_2} + 2\mu_{\ell_2} + {1 \over 3}\mu_{\ell_3}
- \mu'_{d_1} - \mu'_{d_2} - {4 \over 3} \mu'_{e_1} -
{1 \over 3}\mu'_{e_2} + {5 \over 3} \mu'_{\ell_3}
\nonumber \\
{\cal L}'_2 &=&
{1 \over 6} \left(3\mu'_q + 3\mu'_{d_1} + 3\mu'_{d_2}
+ \mu'_{e_1} + \mu'_{e_2} - 17\mu'_{\ell_3}
\right)
\nonumber \\
{\cal L}'_3 &=& \mu'_{e_1}
\nonumber \\
{\cal L}'_4 &=& \mu'_{e_2}
\nonumber \\
{\cal L}'_5 &=& {1 \over 6}\left(
39\mu'_q - 15\mu'_{d_1} - 9\mu'_{d_2} - \mu'_{e_1} - \mu'_{e_2}
- \mu'_{\ell_3}
\right)
\nonumber \\
{\cal L}'_6 &=& 
\mu'_{d_1} - \mu'_{d_2}
\label{a4}
\end{eqnarray}
For the given set of values for the conserved 
quantities, ${\cal L}_i$ and ${\cal L}'_i$, 
the baryon number can be found by solving the identities:
\begin{eqnarray}
B = 12\mu_q
=
12 \times \mu_q + 0 \times \mu_{\ell_2} + 0 \times \mu_{\ell_3} + \ldots
\equiv \alpha_0 {\cal L}_0 + 
\sum_{i=1}^{6} \left( \alpha_i {\cal L}_i + \alpha'_i {\cal L}'_i \right).
\label{a5}
\end{eqnarray}
where it is understood that the ${\cal L}'s$ are also functions
of the 13 independent $\mu$ variables [using Eq.(\ref{a4})].
By equating coefficients of each of the
13 independent variables a set of 13 simultaneous equations
for the $\alpha$'s results:
\begin{eqnarray}
12 &=& 11\alpha_0 + 22\alpha_1 + {1 \over 2}\alpha_2 + {13 \over
2}\alpha_5 - 7\alpha'_1 
\nonumber \\
0 &=& 2\alpha_1 - 4\alpha_0 + 2\alpha'_1
\nonumber \\
0 &=& 2\alpha_1 - {17 \over 6}\alpha_2 - {1 \over 6}\alpha_5
- {1 \over 3}\alpha_0 + {1 \over 3} \alpha'_1
\nonumber \\
0 &=& -\alpha_1 + {1 \over 6}\alpha_2  + \alpha_3 
- {1 \over 6}\alpha_5  - {1 \over 3}\alpha_0 + {1\over 3}\alpha'_1
\nonumber \\
0 &=& {1 \over 6}\alpha_2 + \alpha_4 - {1 \over 6}\alpha_5
- {4 \over 3} \alpha_0 + {1 \over 3}\alpha'_1
\nonumber \\
0 &=& {1 \over 2}\alpha_2 - {5 \over 2}\alpha_5 + \alpha_6
- \alpha_0 + \alpha'_1
\nonumber \\
0 &=& {1 \over 2}\alpha_2 - {3 \over 2}\alpha_5 - \alpha_6
- \alpha_0 + \alpha'_1
\nonumber \\
0 &=& -3\alpha_0 + 29\alpha'_1 + {1 \over 2}\alpha'_2 
+ {13 \over 2}\alpha'_5
\nonumber \\
0 &=& {1 \over 3}\alpha_0 + {5 \over 3}\alpha'_1  - {17 \over
6}\alpha'_2
 - {1 \over 6} \alpha'_5 
\nonumber \\
0 &=& {1 \over 3}\alpha_0 - {4 \over 3}\alpha'_1 + {1 \over 6}\alpha'_2
 + \alpha'_3 - {1 \over 6}\alpha'_ 5
\nonumber \\
 0 &=& {-2 \over 3}\alpha_0  - {1 \over 3}\alpha'_1
 + {1 \over 6}\alpha'_2 + \alpha'_4 - {1 \over 6}\alpha'_5
 \nonumber \\
 0 &=& \alpha_0 - \alpha'_1 + {1\over 2}\alpha'_2 - {5\over 2}\alpha'_5
 + \alpha'_6
 \nonumber \\
 0 &=& \alpha_0 - \alpha'_1 + {1\over 2}\alpha'_2 - {3\over 2}\alpha'_5
 - \alpha'_6
 \end{eqnarray}
These 13 equations can easily be solved for the 13 $\alpha$'s;
the results are as displayed in table \ref{table:case1}.

A similar procedure with $L$ instead of $B$ in Eq. (\ref{a5}) yields 
results for the $\beta$'s.


\begin{thebibliography}{99}

\bibitem{cmb}
D. N. Spergel {\it et al},
(WMAP Collaboration) Astrophys. J. Suppl. 148, 175 (2003),
and references therein for earlier work.


\bibitem{paper1}
R. Foot and R. R. Volkas, Phys.\ Rev.\ D68, 021304 (2003).

\bibitem{bento}
L. Bento and Z. Berezhiani, hep-ph/0111116; see also,
Phys.\ Rev.\ Lett.\ 87, 231304 (2001).

\bibitem{parity}
T. D. Lee and C. N. Yang, Phys.\ Rev.\ 104, 256 (1956);
I. Kobzarev, L. Okun and I. Pomeranchuk, Sov.\ J. Nucl.\ Phys.\ 3,
837 (1966); M. Pavsic, Int.\ J. Theor.\ Phys.\ 9, 229 (1974);
S. I. Blinnikov and M. Yu. Khlopov, Sov. J. Nucl. Phys. 36, 472 (1982);
Sov. Astron. 27, 371 (1983).

\bibitem{parity2}
R. Foot, H. Lew and R. R. Volkas, Phys.\ Lett.\ B272, 67 (1991).

\bibitem{epsilon}
B. Holdom, Phys.\ Lett.\ B166, 196 (1985);
S. L. Glashow, {\it ibid.} B167, 35 (1986);
E. D. Carlson and S. L. Glashow, {\it ibid.} B193, 168 (1987);
R. Foot and X.-G. He, {\it ibid.} B267, 509 (1991);
S. N. Gninenko, {\it ibid.} B326, 317 (1994);
R. Foot and S. Gninenko, {\it ibid.} B480, 171 (2000);
A. Yu. Ignatiev and R. R. Volkas, {\it ibid.} B487, 294 (2000);
R. Foot and R. R. Volkas, {\it ibid.} B 517, 13 (2001);
R. Foot, A. Yu. Ignatiev and R. R. Volkas, {\it ibid.}
B503, 355 (2001);
R. Foot, Acta Phys.\ Polon.\ B32, 3133 (2001);
R. Foot and T. L. Yoon, {\it ibid.} B33, 1979 (2002);
R. Foot and S. Mitra, Phys. Rev. D66, 061301 (2002);
Astropart. Phys. 19, 739 (2003);
Phys. Lett. B558, 9 (2003); Phys. Lett. A315, 178 (2003);
Phys. Rev. D68, 071901 (2003);
Z. Silagadze, astro-ph/0311337;
R. Foot, hep-ph/0308254; astro-ph/0309330.

\bibitem{lss}
Z. Berezhiani, D. Comelli and F. L. Villante, Phys.\ Lett.\ B503, 362 (2001);
A. Yu. Ignatiev and R. R. Volkas, Phys.\ Rev.\ D68, 023518 (2003);
Z. Berezhiani, P. Ciarcelluti, D. Comelli and F. L. Villante,
astro-ph/0312605; P. Ciarcelluti, astro-ph/0312607.

\bibitem{inflation}
E. W. Kolb, D. Seckel and M. S. Turner, Nature (London) 314, 415 (1985);
H. M. Hodges, Phys.\ Rev.\ D47, 456 (1993); 
Z. G. Berezhiani, A. D. Dolgov and R. N. Mohapatra, Phys.\ Lett.\ B375, 26 (1996);
V. Berezinsky and A. Vilenkin, Phys.\ Rev.\ D62, 083512 (2000).

\bibitem{leptogen}
M. Fukugita and T. Yanagida, Phys.\ Lett.\ B174, 45 (1986).

\bibitem{lsnd}
C. Athanassopoulos et al., Phys.\ Rev.\ C54, 2685 (1996);
Phys.\ Rev.\ Lett.\ 81, 1774 (1998); A. Aguilar et al.,
Phys.\ Rev.\ D64, 112007 (2001); see also, G. Drexlin, 
Nucl.\ Phys.\ B (Proc.\ Suppl.) 118, 146 (2003).


\bibitem{mirrornu}
R. Foot, H. Lew and R. R. Volkas, Mod.\ Phys.\ Lett.\ A7, 2567 (1992);
R. Foot, {\it ibid.} A9, 169 (1994);
R. Foot and R. R. Volkas, Phys.\ Rev.\ D52, 6595 (1995).

\bibitem{mohapatra}
R. N. Mohapatra and X. Zhang, Phys.\ Rev.\ D45, 2699 (1992).

\bibitem{kuzmin}
V. A. Kuzmin, V. A. Rubakov and M. E. Shaposhnikov, Phys.\ Lett.\ 155B, 36 (1985).

\bibitem{buch}
See W. Buchm\"{u}ller, hep-ph/0101102 for a review.

\bibitem{shap}
S. Yu.\ Khlebnikov and M. E. Shaposhnikov, Nucl.\ Phys.\ B308, 885 (1988).
There are small higher-order corrections to this result, see
S. Yu.\ Khlebnikov and M. E. Shaposhnikov, Phys.\ Lett.\ B387, 817 (1996);
M. Laine and M. E. Shaposhnikov, Phys.\ Rev.\ D61, 117302 (2000).



\end{thebibliography}
\end{document}